# A Framework for Context-Driven End-to-End QoS Control in Converged Networks

S. Y. Yerima, G.P. Parr, C. Peoples, S. McCLean, P. J. Morrow

*School of Computing and Information Engineering,
University of Ulster, Northern Ireland*
Email: {s.yerima, gp.parr, c.peoples, si.mcclean, pj.morrow}@ulster.ac.uk

*Abstract*— **This paper presents a framework for context-driven policy-based QoS control and end-to-end resource management in converged next generation networks. The Converged Networks QoS Framework (CNQF) is being developed within the IU-ATC project, and comprises distributed functional entities whose instances co-ordinate the converged network infrastructure to facilitate scalable and efficient end-to-end QoS management. The CNQF design leverages aspects of TISPAN, IETF and 3GPP policy-based management architectures whilst also introducing important innovative extensions to support context-aware QoS control in converged networks. The framework architecture is presented and its functionalities and operation in specific application scenarios are described.**

## I. INTRODUCTION

Next Generation Networks (NGNs) will increasingly be characterised by the convergence of fixed and mobile telecommunication systems, Internet and broadcast systems. Convergence of these distinct disparate systems is enabling seamless access to a range of personalised multimedia services through various access technologies such as WiMAX, cellular, xDSL, cable, WiFi, and satellite. Although each of these access technologies may provide mechanisms for Quality of Service (QoS) control, when integrated into a converged NGN environment, unified end-to-end QoS management is important to achieve efficient provision of services with optimal Quality of Experience (QoE) assurance to users anytime, anywhere via any communication device. To this end, this paper proposes a framework for context-aware, end-to-end QoS control designed for highly dynamic converged NGN environments. The Converged Networks QoS Framework (CNQF) provides an infrastructure for end-to-end resource management, QoS provisioning and context-driven adaptation through a policy-based network management (PBNM) paradigm.

Various standards bodies such as the Internet Engineering Task Force (IETF), Third Generation Partnership Project (3GPP), and European Telecommunications Standards Institute's (ETSI) technical committee on Telecoms and Internet converged Services and Protocols for Advanced Networks (TISPAN) have published several PBNM-based architectures which can be found in [1], [2] and [3], for example. While the standards define architectural elements, protocols and interfaces to enable interoperability of different vendor networking equipment, they do not provide details of how end-to-end PBNM solutions may be implemented. Hence, our CNQF design leverages the standard PBNM architectures with important innovative extensions to support context-aware services and adaptation, thus providing a comprehensive infrastructure for end-to-end QoS control and context-driven resource management for next generation converged networks.

Not only is service convergence a reality in today's communication networks, but also heterogeneity of fixed and wireless access methods has become the norm. As such, important requirements of a viable unified end-to-end QoS management solution include flexibility, scalability and extensibility to support the range of fixed and wireless access technologies. Thus, the QoS framework proposed in this paper is designed to fulfill these requirements and comprises distributed functional entities whose instances coordinate the network infrastructure in real-time to facilitate end-to-end QoS control.

Furthermore, with end-user device heterogeneity, mobility and service portability characterising converged networks, it is no longer adequate to assume that the QoS requirements of a session will remain static while provisioning a service. Consequently, adaptation of QoS with respect to network infrastructure as well as service provision is also another essential requirement of end-to-end QoS management. A central feature of our proposed CNQF framework is therefore the ability to support context-aware QoS adaptation mechanisms as well as context-driven resource allocation on an end-to-end basis. The incorporation of context-aware mechanisms within the framework is important because it enables adaptive and intelligent policy-based resource management and QoS control.

The paper continues as follows: Section II gives an overview of related work. Section III describes CNQF and the functionalities of its constituent entities. Section IV discusses the PBNM architecture underpinning CNQF. Some example scenarios supported by CNQF for context-aware QoS control are described in Section V. Lastly, conclusions and future work in the ongoing CNQF project are presented in section VI.

## II. RELATED WORK AND MOTIVATION

As new wired and wireless networking standards and protocols continue to emerge, efforts to address their convergence and management requirements have continued to intensify including research work focusing on frameworks and architectures for QoS management in IP networks. Kim et al., for example, present a QoS framework in [4] based on DiffServe and Simple Network Management Protocol (SNMP) designed to provide QoS control in ad hoc military environments. Similarly, in [5], a policy-based multi-layer QoS architecture for network resource control based on Traffic Engineering (TE) and policy-based routing is presented. Other works addressing policy-based QoS management can be found in [6], [7], [8] and [12], to mention a few. These works however, like majority that can be found in the literature, lack support for context-driven policies or context-aware mechanisms in their solutions; unlike our proposed CNQF where such support is considered an important enabling feature to provide more intelligent and adaptive QoS control.

Although some papers such as [9], [10], and [11] consider context-aware service provisioning, unlike our proposed solution, their studies did not address end-to-end QoS provisioning in converged networks. Our contribution is motivated by the aforementioned limitations in the related approaches. Hence, we propose a policy-based context-driven framework for converged networks' end-to-end QoS control and resource management (CNQF). CNQF is aimed at providing homogeneous, unified, end-to-end QoS support over heterogeneous access technologies, together with scalable, context-aware resource control and adaptive QoS provisioning.

## III. CNQF QoS ARCHITECTURE

CNQF is built around three logical subsystems each consisting of distributed multi-layered entities interacting to enable, monitoring, control and adaptation functions along the end-to-end transport plane. The CNQF subsystems include: a Resource Management Subsystem (RMS), a Context management and Adaptation Subsystem (CAS), and a Measurement and Monitoring Subsystem (MMS). The overall framework is managed by a PBNM system through which high level (and possibly context-aware) policies that govern the behavior of operational entities in the CNQF are specified. The high level declarative policies typically map to lower level operational policies that drive the lower stratum control entities responsible for adaptive configuration of (physical or logical) managed network elements such as routers, gateways, or links.

### A. CNQF Resource Management Subsystem

The RMS is responsible for joint access networks and core networks (CN) resource allocation and control to realise end user QoS provisioning in accordance with Service Level Agreements (SLAs). RMS is made up of distributed instances of peer Resource Brokers (RBs), which will be present in each access network, with each RB maintaining a hierarchical relationship to a Resource Controller (RC) as shown in Fig. 1. An RB will also be present in the core network, in which case it interfaces with and serves as coordinating element to the access network RBs, which could be Fixed Access network Resource Brokers (FARB) or Wireless Access network Resource Brokers (WARB). The core network RB (CNRB) implements policies to manage core network resources whilst the various FARB/WARBs perform a similar role for their respective access networks. In addition to providing CN policy decision functions, CNRB is also responsible for (inter-domain) resource brokerage with external administrative domains.

The RMS enables scalable end-to-end policy-based admission control made possible through interaction between the various WARB/FARBs and the CNRB. All RBs house the policy-driven decision engines which may utilise context information from the Context management and Adaptation Subsystem (CAS) and measurement information from the Measurement and Monitoring Subsystem (MMS) to influence policy decisions/actions. The RBs are the entities that instruct associated RCs to (re)configure network elements with parameters necessary to achieve the QoS control and adaptation goals defined by the high level declarative policies. Thus, RCs are the logical management and control entities that interface with the network elements such as routers, gateways, and access points where policies are enforced. An RC's functions are directed by policies implemented in the associated RB. An RC may be responsible for QoS mapping; for example mapping between Layer 2 QoS scheduling classes in a WiMAX access network (as instructed by a WARB) to IP QoS classes in the CN. Where other access technologies exist in the converged network, an RC is responsible for mapping their respective Layer 2 QoS mechanisms to the Layer 3 IP QoS mechanism adopted in the CN. RCs may also be responsible for packet marking, for example DiffServe Code Point (DSCP) marking in a DiffServe domain. The RC is therefore the entity that controls and co-ordinates directly the low level network mechanisms to realise end-to-end QoS control. Hence, RC is the *policy enforcer* for the RB which serves as the *policy decision point* within the RMS.

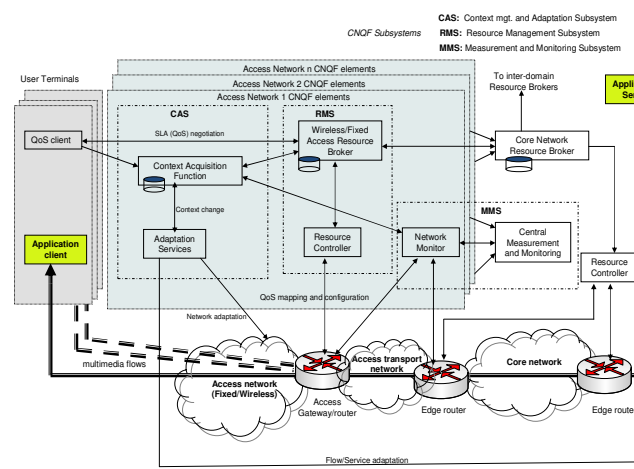

Fig.1. Simplified diagram of CNQF operational entities in a converged network with multiple fixed/wireless access networks.

## B. CNQF Measurement and Monitoring Subsystem

The ability to monitor all devices and network elements is vitally important for the operation of the CNQF. This capability is provided by the Measurement and Monitoring subsystem (MMS) which allows the CNQF to gauge performance, detect faults, and measure SLA compliance within the converged network. The MMS consists of Network Monitors (NM) present in each access network. These interface with a central monitoring and measurement server (CMM) (as shown in Fig. 1) which serves as the aggregating entity for the entire MMS. NMs are the entities responsible for active and passive measurements in the CNQF system. Current implementation of NM passive measurement in the CNQF prototype is based on the Simple Network Management Protocol (SNMP). MMS provides the monitoring and measurement services which are utilized by the RMS and the CAS thus enabling adaptive closed-loop control capabilities within the CNQF policies.

## C. CNQF Context Management and Adaptation Subsystem

An issue of particular importance in converged networks is how *context information* can be exploited to enable adaptive QoS provisioning for a large number of flows. In [13] context is defined as any information, obtained either explicitly or implicitly, that can be used to characterise a certain aspect of an entity involved in a specific application or network service. This entity can be a physical object such as a person, a place, a router, a network gateway, or a physical link; or a virtual object such as an IPsec tunnel or SNMP agent. Clearly, in order to support context-driven QoS control in converged networks, mechanisms for acquisition, processing, storage and retrieval of context-information is necessary. Recognising this, we incorporate context-aware adaptation mechanisms within the CNQF design. This is supported by the Context Management and Adaptation Subsystem (CAS). CAS is composed of distributed Context Acquisition Functions (CAFs) per access network and (distributed or centralised) Adaptation Servers (ADS).

The CAFs provide context acquisition services for the ADS entities and also interface to the RBs in the RMS subsystem to facilitate context-aware resource management. Information such as user preferences, user activities, location, and device capabilities, from a large number of context sources including network monitors, user profiles, etc. are aggregated, processed and used to drive context-aware policies within CNQF. CAF may also infer context from processing other information not directly obtained from context sources. ADS are the entities that undertake reconfiguration of network elements or service parameters (e.g. selection of codecs), under CAF directives. All three subsystems of the CNQF are underpinned by the PBNM architecture described in the next section.

## IV. POLICY-BASED MANAGEMENT INTEGRATION OF CNQF

The requirement for a centralised network management capability in the CNQF framework is addressed by adopting a policy-based management paradigm. This is necessary for deployment of CNQF under an administrative domain otherwise the QoS management of an entire converged network domain would be tedious, time-consuming and error-prone. Policies provide a means to autonomously guide the behavior of a network through high-level declarative directives [12]. The PBNM structure within CNQF is shown in Fig. 2. This structure aligns with policy based management paradigms adopted within IETF [3], TISPAN [2] and 3GPP [1], through distinct separation of policy decision functionality from the policy enforcement points.

The policy management tool provides an interface through which the high-level policies such as context-driven QoS policies, adaptation policies, resource allocation policies, etc. are defined. These are stored in the policy repositories to be retrieved by and processed by the Policy Decision Points (PDPs) when required. The actions to be taken may require enforcement of policies at the Policy Enforcement Points (PEPs) in the transport plane through the various relevant entities like an RC or ADS.

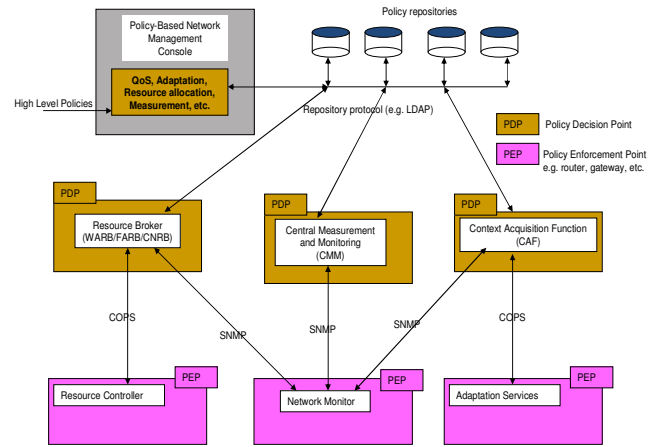

Fig. 2. Policy based network management based integration of CNQF

## V. CONTEXT-AWARE QOS CONTROL SCENARIOS

This section illustrates possible scenarios for context-aware QoS control/adaptation supported by CNQF assuming user mediated session examples. A QoS client may be present in the end user's device to initiate a QoS request during the session set up. Otherwise, the RC may detect the presence of new sessions and notify the relevant WARB/FARB which may accept or reject a session depending on the admission control policies.

From Fig. 3, a user requesting a multimedia service, such as IPTV, signals the QoS requirements to the WARB /FARB via the QoS client. The WARB/FARB will then perform admission control based on end-to-end resource availability which is possible because of its ability to interact with the CNRB. If the policy conditions for the admission are satisfied, the QoS request is granted and a multimedia session is established. A change in 'network context' detected by the CAF would for example trigger a QoS re-configuration request to the associated WARB/FARB which in turn determines the new network configuration parameters with which it instructs the relevant

RC to re-configure the network element(s) accordingly. Session termination could be explicitly notified to the WARB/FARB by the QoS client, in which case the state information is removed and an update sent to the CNRB. Alternatively, the RC may detect inactivity of a session and trigger the same termination process. In this example, the network environment (e.g. Traffic Engineered routes) has been re-configured (in response to congestion) without modifying the transported media flow.

Another variation of the above scenario which involves service adaptation is shown in Fig. 4. In this case, CAS notifies the WARB/FARB of a context change, e.g. session mobility from one device to another (device context), or change in the user location (user context) or activity. WARB/FARB responds to CAS with corresponding QoS parameters after determining resource availability. CAS in turn instructs service adaptation through ADS with new QoS parameters. Upon granting the request, the multimedia flow characteristics are adapted accordingly. Session termination may proceed with similar mechanisms to Fig. 3.

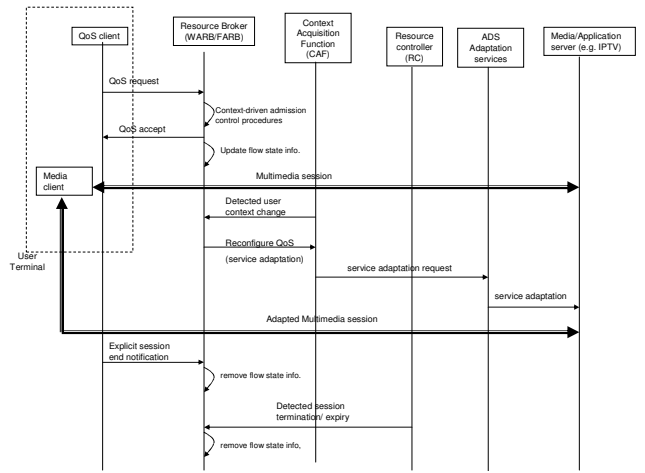

Fig. 4. Procedures for CNQF context-driven service adaptation

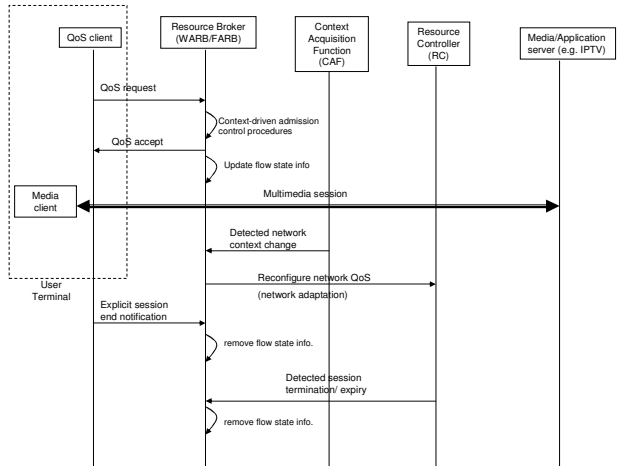

Fig. 3. Procedures for CNQF context-driven QoS adaptation

## VI. CONCLUSION AND FURTHER WORK

This paper presented CNQF, a framework being developed within our project for context-aware policy-based QoS control and resource management in converged next generation networks. In addition to a Resource Management subsystem (RMS) and a Measurement and Monitoring subsystem (MMS), a Context Management and Adaptation subsystem (CAS) is also built into the CNQF design in order to provide context-aware adaptation of QoS and resource management in converged networks. Development of a prototype for validation of CNQF functionalities is currently ongoing within the project. Furthermore, an experimental converged networks testbed is being built within our NETCOM lab at the University of Ulster, for deployment and evaluation of CNQF. Future work will focus on analytical and experimental evaluation of CNQF performance under various use case scenarios using the experimental converged networks testbed.

ACKNOWLEDGMENT

This work is funded by the EPSRC-DST India-UK Advanced Technology Centre of Excellence in Next Generation Networks, Systems and Services (IU-ATC) (www.iu-atc.com) under grant EP/G051674/1.